\documentclass[aps,prb,showpacs,superscriptaddress,twocolumn,floatfix,reprint]
{revtex4-1}

\usepackage{amsmath,amssymb,amsxtra,amstext}
\usepackage{bm}
\usepackage{dcolumn}
\usepackage{graphicx}
\usepackage[utf8]{inputenc}
\usepackage{xspace}

\newcommand\potfit{\textsc{potfit}\xspace}

\begin{document}

\title{Embedded atom method Potentials for Al--Pd--Mn Phases}

\author{Daniel Schopf}
\email{Daniel.Schopf@itap.uni-stuttgart.de}
\affiliation{Institut f\"ur Theoretische und Angewandte Physik (ITAP),
Universit\"at Stuttgart, Pfaffenwaldring 57, 70569 Stuttgart, Germany}
\author{Peter Brommer}
\affiliation{D\'epartement de Physique et Regroupement Qu\'eb\'ecois sur
les Mat\'eriaux de Pointe (RQMP), Universit\'e de Montr\'eal, C.P. 6128,
Succursale Centre-Ville, Montr\'eal, Qu\'ebec, Canada H3C 3J7}
\affiliation{Institut f\"ur Theoretische und Angewandte Physik (ITAP),
Universit\"at Stuttgart, Pfaffenwaldring 57, 70569 Stuttgart, Germany}
\author{Benjamin Frigan}
\author{Hans-Rainer Trebin}
\affiliation{Institut f\"ur Theoretische und Angewandte Physik (ITAP),
Universit\"at Stuttgart, Pfaffenwaldring 57, 70569 Stuttgart, Germany}

\begin{abstract}
A novel embedded atom method (EAM) potential for the $\Xi$ phases of Al-Pd-Mn has
been determined with the force-matching method. Different combinations of analytic
functions were tested for the pair and transfer part. The best results are obtained,
if one allows for oscillations on two different length scales. These potentials
stabilize structure models of the $\Xi$ phases and describe their energy with
high accuracy. Simulations at temperatures up to $1200$ K show very good agreement
with \emph{ab initio} results with respect to stability and dynamics of the system.
\end{abstract}


\maketitle

\section{Introduction}

The ternary intermetallic system Al--Pd--Mn has been of great interest in the last
years, because it forms a high number of complex metallic alloy compounds (CMAs). In
this paper we focus on the $\Xi$ phases, which are approximants of a decagonal
quasicrystal with a lattice constant of $1.6$ nm in the periodic direction. Under
plastic deformation, these phases show a novel type of dislocations, so-called
metadislocations, which were first described by Klein \textit{et al.}\cite{Klein1999}

\emph{Ab initio} studies of these metadislocations, even with fast codes using
density functional theory like VASP,\cite{Kresse1993,Kresse1996} are currently
unfeasible. Their spatial extent is about $200\text{ \AA{}}$ and they involve more
than $10\,000$ atoms -- impossible to simulate even with state of the art
\emph{ab initio} programs.

With classical molecular dynamics (MD) it is easily possible to simulate structures
with millions of atoms in reasonable time. The treatment of atoms as point masses
interacting with an effective potential allows for microscopic insight into
many processes on the atomic scale. The ability to control almost any aspect of the
simulation can be used for optimizing the structure, determining physical properties
or explaining physical phenomena in detail.

However, obtaining an effective potential for classical molecular dynamics is not
straightforward. In order to extract reliable results, a potential has to be
adjusted to the specific physical conditions considered. These can be for example
high pressures, strain, surfaces or phase boundaries. A common way is to fit
a potential such that it reproduces  experimental data like lattice constants,
cohesive and surface energies \cite{Foiles1986,Mei1991} or simply combining pure
element potentials into an alloy potential.

For ternary systems, like Al-Pd-Mn, establishing a potential with these approaches
is very challenging. The small number of available experimental data is not enough
to fit reliable effective potentials. Hence, to obtain a potential that can be used
for structure analysis and optimization, we apply the force-matching method
\cite{Ercolessi1994} using the \potfit package.\cite{Brommer2006,Brommer2007} In the
force-matching method, results from \emph{ab initio} simulations are used as
reference data to adjust the parameters of a potential. This not only dramatically
increases the amount of information available for fitting (the total number of
datapoints can easily reach several thousands). Also, if the reference data is found
to be insufficient, more pertinent reference data can be generated at relatively low
cost. This makes it possible to create realistic potentials for binary or ternary
systems. In our case we used forces on individual atoms, the cohesive energy and
stresses on the unit cells to fit a reliable potential.

In Sec.~\ref{sec:eam} we describe the interaction model used in this research.
The fitting procedure using the force-matching method is presented in
Sec.~\ref{sec:fitting}, the reference data used is given in Sec.~\ref{sec:ref}. The
results will be discussed in detail in Sec.~\ref{sec:results}.

\section{EAM Potentials}
\label{sec:eam}

A common way to describe atomic interactions in metals is the \emph{embedded atom
method} (EAM).\cite{Daw1983} It implicitly includes many-body interactions by a
term which depends on the environment of every atom. The potential energy of a
system described with the EAM method can be written as
\begin{eqnarray}
E_{\text{pot}}=\frac{1}{2}\sum_{\substack{i,j\\j\neq i}}
  \Phi_{ij}(r_{ij})+\sum_iF_i(n_i)\label{eqn:eam1},\\
\text{with}\qquad n_i=\sum_{j\neq i}\rho_j(r_{ij}).
\label{eqn:eam2}
\end{eqnarray}
The first term in \eqref{eqn:eam1} represents the pair interactions between atoms
$i$ and $j$ at a distance $r_{ij}=|\bm{r}_j-\bm{r}_i|$. The function $F_i(n_i)$
is the embedding energy of atom $i$ in the host density $n_i$. This density $n_i$
\eqref{eqn:eam2} is calculated as the sum over contributions from the neighboring
atoms, with $\rho_j$ being the transfer function of atom $j$. It does not represent
an actual physical density; $n_i$ is a purely empirical quantity.

For the pair and transfer part, we have tested three different combinations of
analytic functions as model potentials. Potential I has oscillations in the pair
potential but not in the transfer function. In contrast, potential II has
oscillations only in the transfer function. Finally a third potential has
oscillations in both functions.

For the simple pair potential without oscillations we chose a Morse potential.
It has a single minimum and is used in model II only:
\begin{equation}
\Phi(r)=\Psi\left(\frac{r-r_c}{h}\right)D_e\left[(1-e^{-a(r-r_e)})^2-1\right].
\label{eqn:morse}
\end{equation}
$\Psi$ is a cutoff function, where the free parameters $r_c$ and $h$ describe the
cutoff radius and the smoothing of the potential. The remaining parameters are
$D_e,a,$ and $r_e$; $D_e$ is the depth of the potential minimum, $r_e$ the
equilibrium distance and $a$ the width of the potential minimum. The pair potential
function with oscillations is adopted from Mihalkovi\v{c}
\textit{et al.}:\cite{Mihalkovic2008}
\begin{equation}
\Phi(r)=\Psi\left(\frac{r-r_c}{h}\right)\left[
  \frac{C_1}{r^{\eta_1}}+\frac{C_2}{r^{\eta_2}}\cos(kr+\varphi)\right].
\label{eqn:eopp}
\end{equation}
This ``empirical oscillating pair potential'' (EOPP) has been used in various works
on complex metallic alloys and quasicrystals,
\cite{Mihalkovic1996, Mihalkovic2002, Krajci1992} as it provides great flexibility.
The first term of \eqref{eqn:eopp} with the parameters $C_1$ and $\eta_1$ controls
the short-range repulsion. The second term is responsible for the damping
($C_2,\eta_2$)  of the oscillations with the frequency $k$.

The cutoff function $\Psi(x)$ is defined by
\begin{equation}
\Psi(x)=\frac{x^4}{1+x^4}\\
\label{eqn:cutoff}
\end{equation}
for $x<0$ and $\Psi(x)\equiv0$ for $x\geq 0$. This function guarantees that the
potential functions as well as their derivatives up to the second order approach
zero smoothly at the cutoff distance $r_c$.

Two different analytic forms were used as transfer functions; one allows for
oscillations, the other one does not. The latter one is a simple exponential decay
frequently used in established EAM potentials:\cite{Johnson1989, Pasianot1992,
Mei1991}
\begin{equation}
\rho(r)=\alpha\exp(-\beta r),
\label{eqn:exp_decay}
\end{equation}
where $\alpha$ is the amplitude and $\beta$ is the decay constant. This function
is used in model I. For models II and III, we used an oscillating transfer function,
which is taken from Ref.~\onlinecite{Chantasiriwan1996}:
\begin{equation}
\rho(r)=\Psi\left(\frac{r-r_c}{h}\right)\frac{1+a_1\cos(\alpha r)+a_2\sin(\alpha
r)}{r^\beta}.
\label{eqn:csw}
\end{equation}
The four free parameters are $a_1, a_2, \alpha$ and $\beta$, where $a_1$ and $a_2$
determine the amplitude of the oscillations, $\alpha$ is the wave vector
and $\beta$ controls the decay.

The embedding function $F(n)$ was adopted from Ref.~\onlinecite{Johnson1989}. It is
based on the general equation of state from Rose \textit{et al.}\cite{Rose1984}
The original form is given as
\begin{equation}
F(n)=F_0\left[\frac{q}{q-p}\left(\frac{n}{n_e}\right)^p-
  \frac{p}{q-p}\left(\frac{n}{n_e}\right)^q\right]+F_1\frac{n}{n_e}.
\label{eqn:johnson}
\end{equation}
The parameters in this function are $F_0,F_1,p,q$ and $n_e$. $p$ and $q$ are real
values and $n_e$ is the equilibrium density. In this paper we use this function in
the limit $p\rightarrow q$ and chose $n_e=1$:
\begin{equation}
F(n)=F_0\left[1-q\log n\right]n^q+F_1n,
\label{eqn:pohlong}
\end{equation}
because the original form is numerically unstable with our optimization algorithms.

The number of free parameters of our three potential models is comparatively large.
The non-oscillating (oscillating) pair potential has 3~(6) parameters, and the
non-oscillating (oscillating) transfer function requires 2~(4) values. All models
share the embedding function with 3 free parameters. Every pair and transfer function
has one additional parameter $h$ for the cutoff function $\Psi$. The cutoff radius
$r_c$ is kept fixed at $7\text{ \AA{}}$. In a ternary system like Al--Pd--Mn with 12
potential functions, this adds up to a total number of 60, 48 and 66 parameters for
the models I, II and III, respectively.

\section{Fitting Procedure}
\label{sec:fitting}

All force-matching was performed with the \potfit package of Brommer and Gähler,
\cite{Brommer2006,Brommer2007} which has previously been used to optimize
tabulated pair and EAM potentials. For this work, its capabilities were extended to
analytic potential models.

All free parameters of the analytic functions were fitted to an \emph{ab initio}
reference database containing relaxed ($T=0$) structures, snapshots from
\emph{ab initio} MD simulations at higher temperatures and a few strained samples
(see Tables \ref{tab:structs1} and \ref{tab:structs2}). All \emph{ab initio}
calculations were performed with the Vienna Ab initio Simulation Package (VASP)
\cite{Kresse1993,Kresse1996} using the generalized gradient approximation (GGA) and
the Projector Augmented Wave (PAW) method.\cite{Kresse1999}

Two different optimization algorithms were used to fit the potentials. They both
minimize the sum of squares defined by
\begin{equation}
Z=\sum\omega_E|\Delta E|^2+\sum|\Delta F|^2+\sum\omega_S|\Delta S|^2,
\end{equation}
where $\Delta E$, $\Delta F$ and $\Delta S$ are the energy, force and stress
residuals. These deviations are calculated as the difference of the \emph{ab initio}
and the EAM value, e.g. $$\Delta E=E_{\text{EAM}}-E_{\text{\emph{ab initio}}}.$$
$\omega_E$ and $\omega_S$ are global weights for the energies and stresses.
$\omega_E=22\,500$ was chosen to obtain potentials that yield very precise energies,
but also reasonable forces. For configurations with about 150 atoms, this effectively
weighs the energies with a factor of approximately 50. The stress weight $\omega_S$
was set to 750, so that the total weight of the six stress tensor components per
configuration is approximately equal to ten times the weight of all forces in one
configuration.

The first optimization algorithm used is simulated annealing.\cite{Kirkpatrick1983}
It is based on the Metropolis criterion, where a decrease in the target function $Z$
is always accepted and an increase only with a probability $P=e^{-\Delta Z/T}$. This
allows the algorithm to escape local minima. The artificial temperature $T$ is
steadily decreased during the optimization. To ensure that the fit converged to the
global minimum, the optimization was restarted with a high temperature several times.
Subsequently a conjugate gradient based method \cite{Powell1965} was applied to
converge to the final optimum. During the fitting procedure, all parameters were
confined to a predefined range by use of numerical punishments.

\section{Reference data}
\label{sec:ref}

The structures used as reference data are shown in Tables \ref{tab:structs1} and
\ref{tab:structs2}. There are 119 configurations with a total of $16\,103$ atoms.
The number of reference datapoints is $49\,340$. They consist of $48\,309$ forces,
119 energies and 714 stresses.

\begin{table}[htp]
\begin{ruledtabular}
\begin{tabular}{cccc}
& Al--Mn structures & Al--Pd structures & \\ \hline
& Al$_{10}$Mn$_3$.\textit{hP}26 & AlPd.\textit{cP}8 & \\
& Al$_{11}$Mn$_4$.\textit{aP}15  & Al$_{21}$Pd$_8$.\textit{tI}116 & \\
& Al$_{12}$Mn.\textit{cI}26  & Al$_3$Pd$_2$.\textit{hP}5 & \\
& Al$_6$Mn.\textit{oC}28 & & \\
& AlMn.\textit{tP}4 & &
\end{tabular}
\end{ruledtabular}
\caption{Binary structures ($T=0$) used to fit the potentials, with their
corresponding Pearson symbol.}
\label{tab:structs1}
\end{table}

In addition to the binary and ternary structures, one reference configuration
for each of the pure elements was also included. These were, in detail,
Al.\textit{cF}4, Pd.\textit{cF}4 and Mn.\textit{cI}58. This was done to get
reliable reference points for the calculation of the enthalpy of formation.

\begin{table}[htp]
\begin{ruledtabular}
\begin{tabular}{lcc}
 & Number of atoms & $\Delta H$ (eV/atom) \\ \hline
$T=0$ & Al$_{92}$Pd$_{28}$Mn$_{10}$\footnotemark[1] & $-0.512$ \\
 & Al$_{92}$Pd$_{28}$Mn$_8$\footnotemark[1] & $-0.485$ \\
 & Al$_{112}$Pd$_{36}$Mn$_6$\footnotemark[2] & $-0.526$ \\
 & Al$_{114}$Pd$_{34}$Mn$_6$\footnotemark[2] & $-0.503$ \\
 & Al$_{112}$Pd$_{34}$Mn$_6$\footnotemark[2] & $-0.512$ \\
 & Al$_{110+x}$Pd$_{32}$Mn$_{8}$\footnotemark[2] & see Sec. \ref{subsec:refinement}
\\
 & Al$_{124}$Pd$_{8}$Mn$_{24}$\footnotemark[2]\footnotemark[3] & $-0.297$ \\
 & Al$_{147}$Pd$_{43}$Mn$_{18}$\footnotemark[2] & $-0.485$ \\
& Al$_{294}$Pd$_{88}$Mn$_{16}$\footnotemark[2] & $-0.491$ \\
$T>0$ & Al$_{92}$Pd$_{28}$Mn$_8$\footnotemark[1]\footnotemark[4] & -- \\
 & Al$_{92}$Pd$_{28}$Mn$_{10}$\footnotemark[1] ($1500$ K) & -- \\
\end{tabular}
\end{ruledtabular}
\footnotetext[1]{Structure generated from canonical cell tiling \cite{Henley1991}}
\footnotetext[2]{From structure optimization}
\footnotetext[3]{T-Al-Pd-Mn, see Sec.~\ref{sec:ref}}
\footnotetext[4]{From several MD runs at $600$, $1100$ and $1800$ K with small
strains}
\caption{Ternary structures used to fit the potentials and their \emph{ab initio}
formation enthalpy $\Delta H$.}
\label{tab:structs2}
\end{table}

All atomic configurations from binary systems (Table~\ref{tab:structs1}) were
taken from the alloy database of Widom \textit{et al.}\cite{alloydb} and have
been fully relaxed with \emph{ab initio} methods. They were chosen to provide more
data for the Pd--Pd and Mn--Mn interactions. Magnetism was not included in
our \emph{ab initio} calculation; it was shown that the manganese atoms in the
$\Xi$-phases are nonmagnetic.\cite{Hippert1999} Because the structures we
want to investigate are on the aluminum-rich side of the phase diagram,
there is only little data for the Mn--Pd interaction.

The reference configurations for the $\Xi$-phases are from different sources.
The structures in Table~\ref{tab:structs2} denoted with superscript a were taken
from the alloy database.\cite{alloydb} They were generated with the canonical cell
tiling,\cite{Henley1991} which creates hypothetical models by decorating a tiling
with clusters. To compensate for the low amount of manganese in these samples
and the hence resulting lack of data, five of the aluminum atoms were replaced
by manganese in some configurations. \emph{Ab initio} molecular dynamics simulations
with VASP \cite{Kresse1993,Kresse1996} were run with these samples at $600$, $1100$
and $1800$ K to obtain different local atomic configurations. These calculations
were done in the generalized gradient approximation (GGA) with PAW
potentials.\cite{Kresse1999}

At the same time, \emph{ab initio} structure optimization was carried out for two of
the $\Xi$-phases. Particularly this were the $\Xi$-phase with the smallest unit cell,
which contains about 152 atoms and is called $\xi$ and the next bigger one,
containing about 304 atoms, which is called $\xi'$. All structures generated in the
course of this optimization are denoted in Table \ref{tab:structs2} by superscript b.

To judge the stability of these structures, their energy is compared to a mixture
of competing phases, the convex hull. This hull, defined over a ternary phase
diagram, contains the cohesive energies of all stable compounds as vertices.
If the energy of a structure is above this convex hull, it could decompose
into the neighboring  structures and thus lower its energy. If the energy
of a new structure is below the convex hull, it is considered to be
thermodynamically stable. The structures which define the convex hull for the
$\Xi$-phases, are T-AlPdMn, Al$_{12}$Mn, Al$_{21}$Pd$_8$ and Al$_3$Pd$_2$. They
have also been included in the reference database. A detailed description of these
phases and the convex hull is given in Ref.~\onlinecite{Frigan2011}.

\section{Results}
\label{sec:results}

We determined parameters for all three potential models from the reference data
described above. The root mean square (RMS) errors for forces, energies and
stresses after the optimization are in the same order of magnitude for all
models (see Table \ref{tab:rms_opt}). While model III has the smallest errors for
forces and energies, model I has the biggest errors for all three quantities.
Model II has the smallest stress deviations. While the force error for model I is
about 20\% larger than the one for model III, the energy error is significantly
larger with about 50\% difference.

\begin{table}[htp]
\begin{ruledtabular}
\begin{tabular}{lrrr}
RMS errors for & \multicolumn{1}{c}{Model I} & \multicolumn{1}{c}{Model II} &
  \multicolumn{1}{c}{Model III} \\ \hline
forces & $265.63$ & $221.40$ & $220.07$ \\
energies & $19.36$ & $14.49$ & $12.53$ \\
stresses & $99.99$ & $76.83$ & $98.30$
\end{tabular}
\end{ruledtabular}
\caption{Root mean square errors after the optimization for forces (in meV/\AA),
energies (in meV/atom) and stresses (in kPa). This data is calculated
with the reference configurations used for fitting the potentials.}
\label{tab:rms_opt}
\end{table}

\begin{figure}[htp]
\includegraphics{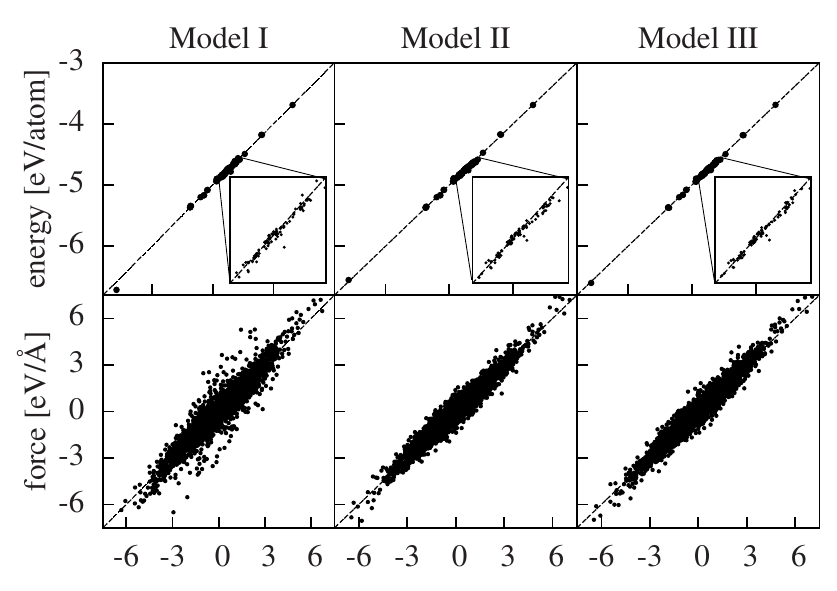}
\caption{Scatterplot for energies and forces with the EAM values on the vertical
axis and the \emph{ab initio} reference data on the horizontal axis. The insets are
magnified by a factor 4.5.}
\label{fig:scatter}
\end{figure}

A graphical representation of these errors can be seen in Fig.~\ref{fig:scatter}.
The scatterplots in the upper row display the energies of the reference data.
Forces are shown in the lower row. The range of the force plots is due to the many
high temperature MD simulations that are included in the reference data. The forces
therein can become very large because of the short interatomic distances that may
occur at these temperatures.

These errors cannot solely be used to judge the quality and transferability of the
potentials. For that purpose another set of \emph{ab initio} data has been extracted
from
the structure optimization. It has not been included in the reference data and
can be used to determine the transferability of the different potentials.
The same errors as before have been calculated and can be seen in Table
\ref{tab:rms_test}. As with the reference data, model III has the lowest force
and energy errors. The relative error of the energy is about 0.2\%, for stresses
about 5\% and 550\% for forces. This is due to the fact that all configurations in
this test data are ground state structures and therefore only contain very small
forces.

\begin{table}[htp]
\begin{ruledtabular}
\begin{tabular}{lrrr}
RMS errors for & \multicolumn{1}{c}{Model I} & \multicolumn{1}{c}{Model II} &
  \multicolumn{1}{c}{Model III} \\ \hline
forces & $141.90$ & $131.90$ & $130.46$ \\
energies & $10.42$ & $10.47$ & $10.28$ \\
stresses & $32.39$ & $23.76$ & $36.89$ \\
\end{tabular}
\end{ruledtabular}
\caption{Root mean square errors for forces (in meV/\AA), energies (in meV/atom) and
stresses (in kPa). This data is calculated with test data, containing
only structures that were not included in the optimization process.}
\label{tab:rms_test}
\end{table}

The errors for the test data in Table \ref{tab:rms_test} are smaller than those of
the reference configurations in Table \ref{tab:rms_opt}, because there are only
ground states included and no high temperature MD runs.

Based upon these simple energy and force considerations, all the potential models
appear to be of similar quality. Model III, however, should be slightly superior to
the other two potentials. Further tests are necessary to determine the
performance of the potentials in different situations. They will be presented in
Subsection~\ref{subsec:tests}.

\subsection{Structure Refinement}
\label{subsec:refinement}

In Ref.~\onlinecite{Frigan2011}, the structure of the $\Xi$-phases of
Al--Pd--Mn has been optimized by energy minimization in \emph{ab initio} and
molecular dynamics simulations. We use several of the structures tested there to
judge the quality of the optimized potentials. The $\Xi$-phases consist of
columns of pseudo-Mackay icosahedral clusters (PMIs),\cite{Sun1996} a slight
deviation of the famous Mackay icosahedron.\cite{Mackay1962}

\begin{figure}[htp]
\begin{center}
\includegraphics{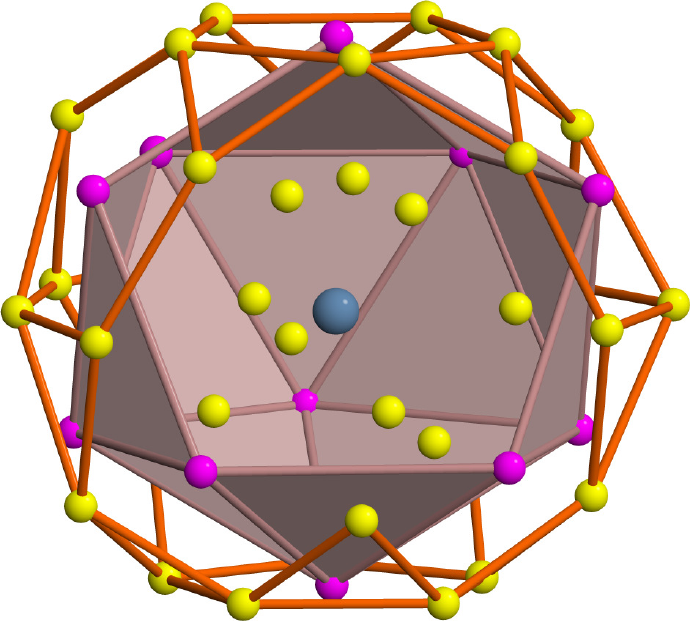}
\caption{(color online) Detailed structure of the pseudo-Mackay icosahedral cluster.
A few atoms of the second and third shell are omitted to see the central atom and
the aluminum atoms in the first shell. The icosahedron and icosidodecahedron are
indicated by planes and bars, respectively. The central atom is manganese. Aluminum
atoms are depicted in yellow (light gray), and palladium atoms in magenta (dark
gray).}
\label{fig:pmi}
\end{center}
\end{figure}

Every PMI cluster consists of a single atom at the center with a first shell of an
experimentally poorly determined number of aluminum atoms. The second shell is an
icosahedron of 12 transition metal atoms and the outer shell an icosidodecahedron of
30 aluminum atoms, see Fig.~\ref{fig:pmi}. Almost all atoms of the $\Xi$-phases
belong to these clusters. It is difficult to measure the exact number of atoms in the
first shell because aluminum atoms are hard to observe in diffraction experiments.

For the lowest quasicrystal approximant, the $\xi$-phase, there are four PMI clusters
in one unit cell. Several different occupancies of aluminum atoms in the first
shell were tested in Ref.~\onlinecite{Frigan2011}. Each configuration was
denoted by a single number, giving the average number of aluminum atoms per
cluster. Structures from eight up to eleven atoms per PMI were generated and tested.

The results with the different potential models can be seen in
Table~\ref{tab:engdiff}. All structures were completely relaxed with \emph{ab initio}
methods, the corresponding \emph{ab initio} energy is given in column 2. The energies
of these configurations with the generated EAM potentials have been calculated after
subsequent relaxation with the respective potentials. This relaxation causes small
displacements of the atoms from their \emph{ab initio} determined positions. For
models I these average displacements are 0.10~\AA{}/atom, 0.08~\AA{}/atom for model
II and 0.11~\AA/atom for model III. This clearly shows that all potential
models can stabilize the ground states of all structures that were generated.

\begin{table}[htp]
\begin{ruledtabular}
\begin{tabular}{D{.}{.}{3}rrrr}
\multicolumn{1}{c}{Number of atoms} & \multicolumn{1}{c}{$E_\text{\emph{ab
initio}}$} &
\multicolumn{3}{c}{$\Delta E$ (meV/atom)}\\
\multicolumn{1}{c}{per PMI} & (eV/atom) & Model I & Model II & Model III \\ \hline
8 & $-4.753$ & $-13$ & $-12$ & $-20$\\
8.25 & $-4.755$ & $-6$ & $-7$ & $-13$\\
8.5 & $-4.756$ & $-1$ & $-3$ & $-5$\\
8.75 & $-4.757$ & $+3$ & $+2$ & $+2$ \\
9 & $-4.755$ & $+4$ & $+4$ & $+3$\\
9.25 & $-4.747$ & $+1$ & $+1$ & $+1$ \\
9.5 & $-4.741$ & $+1$ & $+0$ & $+2$\\
9.75 & $-4.731$ & $-6$ & $-5$ & $-2$\\
10 & $-4.731$ & $0$ & $+2$ & $+4$\\
10.25 & $-4.714$ & $-12$ & $-12$ & $-5$\\
10.5 & $-4.704$ & $-15$ & $-17$ & $-7$\\
10.75 & $-4.692$ & $-19$ & $-21$ & $-13$\\
11 & $-4.683$ & $-22$ & $-24$ & $-17$\\
\end{tabular}
\end{ruledtabular}
\caption{Cohesive energies (in eV/atom) of different optimized
configurations for the $\xi$-phase. The energy differences $\Delta E$ between the
\emph{ab initio} calculations and the respective model are given in meV/atom.}
\label{tab:engdiff}
\end{table}

All models are having difficulties with the energies of structures that contain less
than 9 or more than 10 atoms in the inner shells of the PMI clusters. This may be an
indication for the mechanical instability found during the structure
optimization.\cite{Frigan2011} The energy of these structures is highly unfavorable;
at elevated temperatures some atoms drifted from the outer shell to the inner shell
or vice versa to achieve an inner shell with 9 or 10 aluminum atoms.

All energy differences between the \emph{ab initio} and EAM calculations are smaller
than 10 meV/atom for configurations ranging from 8.5 to 10 atoms per
PMI cluster. This energy is considered a critical threshold for the accuracy of the
potentials. Regarding the energy differences between the different structures, which
are on the order of 1 meV/atom, all potentials can evidently distinguish between
these different configurations.

The structure optimization in Ref.~\onlinecite{Frigan2011} yielded four almost
stable structures, which are different from the ones shown in
Table~\ref{tab:engdiff}. There, not only the atoms in the inner shell are varied, but
also atoms not belonging to the PMI clusters. These alterations were not done in a
systematic manner, the structures will be listed in tabular form. The amount of
atoms for $\xi$- and $\xi'$-phases is the same, only the arrangement of the PMI
cluster columns is different. These structures were tested with the three different
potentials. The results can be seen in Table \ref{tab:engdiff_stable}. The two upper
structures in this table are $\xi$-phase, the lower two structures are $\xi'$.

\begin{table}[htp]
\begin{ruledtabular}
\begin{tabular}{rrrrr}
& \multicolumn{1}{c}{$E_\text{\emph{ab initio}}$} &
\multicolumn{3}{c}{$\Delta E$ (meV/atom)}\\
composition & (eV/atom) & Model I & Model II & Model III \\ \hline
$\xi$-228--64--12 & $-4.702$ & $-5$ & $+4$ & $0$  \\
$\xi$-224--68--12 & $-4.748$ & $+1$ & $+7$ & $+4$  \\
$\xi'$-228--64--12 & $-4.703$ & $-5$ & $+3$ & $+1$  \\
$\xi'$-224--68--12 & $-4.748$ & $+1$ & $+5$ & $+5$
\end{tabular}
\end{ruledtabular}
\caption{Cohesive energies (in eV/atom) of the four almost stable phases after
relaxation. The energy differences $\Delta E$ are given in meV/atom. The composition
is given in numbers of aluminum, palladium and manganese atoms, in this order. All
configurations have 9 aluminum atoms in the inner shell of the PMI clusters.}
\label{tab:engdiff_stable}
\end{table}

After the relaxation with the effective potentials, all models show a very good
agreement with the \emph{ab initio} calculated energies. The mean displacements after
the relaxation are again in the same order of magnitude as before, 0.11~\AA{}/atom
for model I, 0.08~\AA{}/atom for model II and 0.15~\AA{}/atom for model III.

Based on these pure energy comparisons, all three potential models seem to be of
equal quality, with slight advantages for model III.

\subsection{Tests}
\label{subsec:tests}

A force-matched potential is only useful, if it can reproduce key quantities
that were not directly included in the reference data. Here, we subjected the
three potentials to a series of tests. The first test is whether the potential
can stabilize the $\xi$-phase even at elevated temperatures. As there was a
large number of high temperature \emph{ab initio} MD simulations included in the
optimization, the potentials should be able to preserve the structure of the
$\xi$-phase under these conditions. We carried out an \emph{ab initio} MD simulation
at 1200 K for 50 ps,\cite{Frigan2011} where the phase is still mechanically stable.
In a time-averaged picture of the density, the atoms in the two outer shells of the
PMI clusters did not move, but the atoms in the first shell showed some rotational
degree of freedom.

All three models were able to stabilize the structure at this temperature. While
models II and III give the same results as the \emph{ab initio} calculation (cf.\
Ref.~\onlinecite{Frigan2011}), model I shows additional degrees of freedom. In the
time-averaged picture the atoms forming the outer shell of the PMIs are not as
steady as in the \emph{ab initio} simulation. Also the atoms, which do not belong the
these clusters, exhibit a density distribution that is twice as large as expected.
This means that model I may have difficulties stabilizing the structure at even
higher temperatures or against fluctuations in the local atomic arrangement.

For molecular dynamics simulations, the stabilization of different phases can be a
problem. We checked some well known phases for all three potential models
with respect to cohesive energy and phase stability. The results can be seen in
Table \ref{tab:suspects}.
All three potentials can stabilize the different phases. The deviation of the atomic
positions after relaxation compared to the \textit{ab initio} reference values is
very small. The energies are reproduced with errors of under 200 meV/atom.

\begin{table*}[htp]
\begin{ruledtabular}
\begin{tabular}{ccrrrrrrrrr}
 & \multicolumn{1}{c}{$E_\text{\emph{ab initio}}$} &
\multicolumn{3}{c}{Model I} &
\multicolumn{3}{c}{Model II} & \multicolumn{3}{c}{Model III} \\
System & $E$ (eV/atom) & $E_{\text{EAM}}$ & $\Delta E$ & $\Delta x$ &
$E_{\text{EAM}}$ & $\Delta E$ & $\Delta x$ & $E_{\text{EAM}}$ & $\Delta E$ &
$\Delta x$ \\ \hline
AlPd.\textit{cP}2 (B2) & -5.330 & -5.430 & -0.100 & 0 & -5.503 & -0.173 & 0 & -5.445
& -0.115
& 0 \\
AlPd$_3$.\textit{cF}16 (D$0_{3}$) & -5.236 & -5.426 & -0.190 & 0 & -5.424 & -0.188 &
0 &
-5.430 & -0.194 & 0 \\
Al$_3$Pd.\textit{tI}8 (D$0_{22}$) & -4.421 & -4.546 & -0.125 & 0 & -4.540 & -0.119 &
0 &
-4.560 & -0.139 & 0\\
Al$_3$Pd.\textit{cP}4 (L1$_{2}$) & -4.609 & -4.647 & -0.038 & 0.13 & -4.651 & -0.042
& 0.17
& -4.650 & -0.041 & 0.17 \\ \hline
Al$_3$Mn.\textit{tI}8 (D$0_{22}$) &  -5.132 & -5.053 & 0.079 & 0.04 & -5.129 & 0.003
& 0.01
& -5.175 & -0.043 & 0\\
Al$_3$Mn.\textit{cP}4 (L1$_{2}$) & -5.032 & -5.187 & -0.155 & 0 & -5.180 & -0.148 & 0
& -5.197
& -0.165 & 0
\end{tabular}
\end{ruledtabular}
\caption{Cohesive energies for different phases in the Al-Pd-Mn system. All energies
and energy differences are given in eV/atom. The mean square displacements ($\Delta
x$) after relaxation are given in \AA{}/atom. A displacement of 0 means the value is
smaller than 10$^{-4}$ \AA{}/atom.}
\label{tab:suspects}
\end{table*}

\begin{figure}[htp]
\begin{center}
\includegraphics{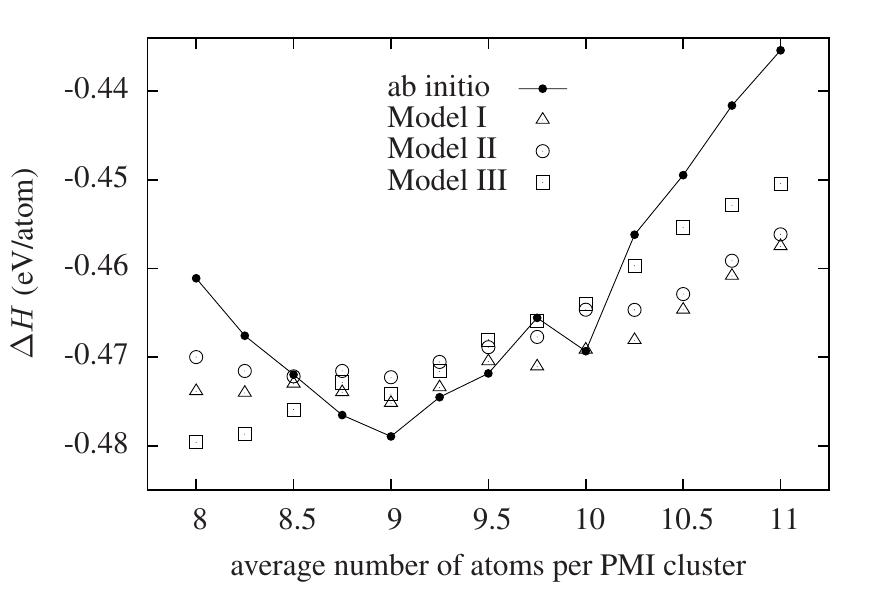}
\caption{Comparison of the \emph{ab initio} formation enthalpy $\Delta H$ (in
eV/atom) with the three potential models. The lines between the \emph{ab initio}
datapoints are added as a guide to the eye.}
\label{fig:enthalpy}
\end{center}
\end{figure}

Another important test is the calculation of formation enthalpies $\Delta H$ with
the potentials.
$\Delta H$ is defined as the energy difference of a structure to the tie plane
of the pure element energies. This has been calculated for all configurations in
Tables \ref{tab:engdiff} and \ref{tab:engdiff_stable}. The reference energies are
given in Table \ref{tab:pureelements}. For the structures with different amounts of
aluminum atoms in the inner shell of the PMI clusters, the results can be seen in
Figure~\ref{fig:enthalpy}. The deviations from the \emph{ab initio} enthalpies are
very similar to those from Table~\ref{tab:engdiff}. For less than 8.5 and more than
10 atoms in the inner shell of the PMI clusters the enthalpies differ more than 10
meV/atom.

\begin{table}[htp]
\begin{ruledtabular}
\begin{tabular}{crrrr}
 & $E_\text{\emph{ab initio}}$ &
\multicolumn{3}{c}{$\Delta E$ (meV/atom)} \\
 & (eV/atom) & Model I & Model II & Model III \\ \hline
Al.\textit{cF}4 & $-3.688$ & $-5$ & $-4$ & $-3$ \\
Pd.\textit{cF}4 & $-5.199$ & $0$ & $0$ & $0$ \\
Mn.\textit{cI}58 & $-8.964$ & $+68$ & $-7$ & $+3$ \\
\end{tabular}
\end{ruledtabular}
\caption{\emph{Ab initio} energies (in eV/atom) and the differences for the
effective potentials (in meV/atom) for the pure elements. These energies were
used for the calculation of the formation enthalpies $\Delta H$.}
\label{tab:pureelements}
\end{table}

The enthalpies for the four almost stable structures are shown in Table
\ref{tab:formationenthalpy_stable}. All three models give very accurate enthalpies
with deviations all smaller than 10 meV/atom.

\begin{table}[htp]
\begin{ruledtabular}
\begin{tabular}{rrrrr}
 & \multicolumn{1}{c}{$\Delta H_{\text{\emph{ab initio}}}$} &
\multicolumn{3}{c}{$\Delta H_{\text{EAM}}-\Delta H_{\text{\emph{ab initio}}}$}\\
composition & (eV/atom) & Model I & Model II & Model III \\ \hline
$\xi$-228--64--12 & $-0.488$ & $-6$ & $+1$ & $-1$ \\
$\xi$-224--68--12 & $-0.513$ & $0$ & $+4$ & $+3$ \\
$\xi'$-228--64--12 & $-0.488$ & $-6$ & $0$ & $0$ \\
$\xi'$-224--68--12 & $-0.514$ & $0$ & $+2$ & $+3$ \\
\end{tabular}
\end{ruledtabular}
\caption{\emph{Ab initio} formation enthalpies $\Delta H$ (in eV/atom) of the four
almost stable phases and the differences for the effective potentials (in meV/atom)
after relaxation. The composition is given in numbers of aluminum, palladium and
manganese atoms, in this order. All configurations have 9 aluminum atoms inside the
PMI clusters.}
\label{tab:formationenthalpy_stable}
\end{table}

During the structure optimization a very long \emph{ab initio} MD run with
$50\,000$ steps at 1200 K was performed. Snapshots were taken from this
simulation at different timesteps and quenched very rapidly. This has also
been done with the EAM potentials. The results show a very good agreement
for different snapshots. The structures only differ very slightly in atomic
positions. While there is a steady offset of about 100 meV/atom in the
energy for higher temperatures, the overall trend can clearly be followed.
For lower temperatures and $T=0$ the energies were in the same order as for
the structures in Table \ref{tab:engdiff}. There were no major
differences for all three potential models.

\begin{figure}[htp]
\begin{center}
\includegraphics{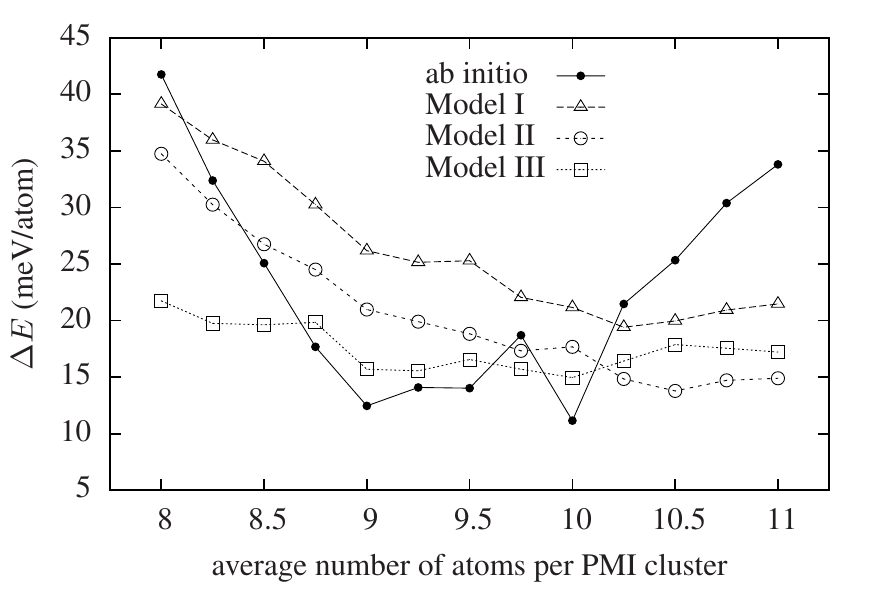}
\caption{Difference of the cohesive energy to the convex hull for different amounts
of aluminum atoms in the inner shell of the PMI clusters. The datapoints for model
I are shifted by +10 meV/atom and by +5 meV/atom for model II.}
\label{fig:convexhull}
\end{center}
\end{figure}

To determine if a structure is thermodynamically stable, the energy difference of
this structure to the convex hull is calculated. If this difference is negative, the
structure is stable, otherwise it it unstable. For more details on the convex hull
see Ref.~\onlinecite{Frigan2011}. This energy difference has been calculated for all
structures in Table~\ref{tab:engdiff} and is shown in Fig.~\ref{fig:convexhull}.

For the sake of clarity, the datapoints of model I and II are shifted by +10
and +5 meV/atom. While these models show a clear decrease of the energy difference
with increasing number of atoms inside the PMI cluster, model III has minima for 9
and 10 atoms, like the \emph{ab initio} reference calculation. As this is the main
criterion for performing a structure optimization, models I and II cannot be used
for this purpose. Only model III is able to reproduce the shape of
the \emph{ab initio} calculation.

The melting point for the $\xi$-phase has been determined with all three potential
models. In MD simulations the volume per atom has been calculated while the sample
was heated from 950 K to 1400 K. At the melting point, there is a jump in atomic
volume, which corresponds to the melting transition. For model I this was found at
1130 K, for model II at 1370 K and for model III at 1300 K. With this method, the
melting point is generally overestimated, due to the high heating rates. For the
simulations we chose a heating rate of $5\times 10^{-5}$ K per timestep, this equals
$5\times 10^9$ K/s. If one compares these temperatures with the experimental value of
1118 K, the value for the potential model I seem to be too low, model II and III are
in the expected temperature range.

Another test we performed is the calculation of the elastic constants.
All $\Xi$-phases have an orthorhombic unit cell. The corresponding nine
elastic constants were determined by examining the cohesive energy during
homogeneous deformations of the sample.

\begin{table}[htp]
\begin{ruledtabular}
\begin{tabular}{crrrr}
-- & \emph{ab initio} & Model I & Model II & Model III \\ \hline
$C_{11}$ & $175.79$ & $255.25$ & $244.66$ & $200.98$ \\
$C_{22}$ & $192.75$ & $269.79$ & $246.74$ & $193.61$ \\
$C_{33}$ & $227.46$ & $243.53$ & $246.64$ & $160.57$ \\
$C_{12}$ & $58.76$ & $158.83$ & $145.57$ & $102.76$ \\
$C_{13}$ & $67.85$ & $146.75$ & $146.78$ & $92.95$ \\
$C_{23}$ & $56.34$ & $151.19$ & $146.51$ & $107.04$ \\
$C_{44}$ & $72.54$ & $42.57$ & $42.42$ & $42.77$ \\
$C_{55}$ & $67.77$ & $41.46$ & $47.19$ & $46.66$ \\
$C_{66}$ & $71.25$ & $48.51$ & $48.21$ & $43.76$ \\
\end{tabular}
\end{ruledtabular}
\caption{Elastic constants of $\xi$-Al--Pd--Mn in GPa.}
\label{tab:elconst}
\end{table}

The results with all three models (see Table \ref{tab:elconst}) show only very little
agreement with the \emph{ab-initio} values. Only model III can reproduce $C_{11}$ and
$C_{22}$. All other elastic constants differ by up to a factor of 3. The potentials
are apparently not able to reproduce the shear stress. However, this behavior is to
be expected, if one takes into account that these potentials were generated for
energy minimization purposes. For other applications, like calculating mechanical
properties, an extended database, containing enough data on shears, should be used.
The only samples used for these potentials, that included deformations, were high
temperature \emph{ab-initio} MD snapshots. These were strained along either of the
cartesian axes, which are perpendicular to the periodic stacking axis of the
quasicrystal. The corresponding elastic constants are $C_{11}$ and $C_{22}$, which
are the only ones correctly reproduced by model III.

This clearly shows that force matched potentials are limited in their applications.
They give very accurate results regarding the energy and forces because they are
tuned to these quantities. For other physical properties, like elastic
constants, the potentials are less accurate.

\section{Summary}

The Al--Pd--Mn potentials presented are very well suited to model the
energetics of the $\Xi$-phases. They were obtained with the force-matching
method, which is fitting the parameters to a large database of \emph{ab initio}
determined reference data. All three analytic potential models tested were
able to reproduce the \emph{ab initio} values of the energies with very high
accuracy. The error sum of the fitting process for all three potentials is very
similar, yet they show very different properties when used in MD simulations.

\begin{table*}[htp]
\begin{ruledtabular}
\begin{tabular}{crrrrrrr}
\multicolumn{8}{c}{EOPP pair function} \\
pair & \multicolumn{1}{c}{$C_1$} & \multicolumn{1}{c}{$\eta_1$}
  & \multicolumn{1}{c}{$C_2$} & \multicolumn{1}{c}{$\eta_2$}
  & \multicolumn{1}{c}{$k$} & \multicolumn{1}{c}{$\varphi$}
  & \multicolumn{1}{c}{$h$}\\ \hline
Al-Al & $586.4805$ & $7.6769$ & $-0.0333$ & $1.0012$ & $3.7658$ & $3.8484$
  & $1.3897$ \\
Al-Mn & $338.7250$ & $7.5484$ & $-0.4212$ & $1.9271$ & $2.7530$ & $0.0033$
  & $0.5000$ \\
Al-Pd & $981.8107$ & $9.1908$ & $-89.9193$ & $4.7322$ & $0.2491$ & $1.3235$
  & $0.6211$ \\
Mn-Mn & $3.8460$ & $19.9995$ & $-44.5953$ & $4.1469$ & $1.2084$ & $1.0115$
  & $1.5938$ \\
Mn-Pd & $12.8931$ & $3.4348$ & $-90.3824$ & $4.4851$ & $1.6212$ & $0.0005$
  & $0.5007$ \\
Pd-Pd & $6625.3081$ & $9.5962$ & $99.8792$ & $6.1164$ & $3.8088$ & $2.5086$
  & $0.5235$ \\
\hline \multicolumn{7}{c}{transfer function} \\
element & \multicolumn{1}{c}{$a_1$} & \multicolumn{1}{c}{$a_2$}
  & \multicolumn{1}{c}{$\alpha$} & \multicolumn{1}{c}{$\beta$}
  & \multicolumn{1}{c}{$h$} \\ \hline
Al & $0.1317$ & $0.0399$ & $2.7507$ & $2.3142$ & $1.9995$ \\
Mn & $-1.5432$ & $1.0321$ & $1.6018$ & $2.4154$ & $1.9996$ \\
Pd & $0.4962$ & $0.7317$ & $2.9972$ & $3.4308$ & $0.5001$ \\
\hline \multicolumn{7}{c}{embedding function} \\
element & \multicolumn{1}{c}{$F_0$} & \multicolumn{1}{c}{$F_1$}
  & \multicolumn{1}{c}{$q$} & & & \\ \hline
Al & $-2.9403$ & $0.5639$ & $-1.3026$ \\
Mn & $-1.5862$ & $1.3917$ & $-5.3935$ \\
Pd & $-4.0016$ & $0.9432$ & $-5.7749$ \\
\hline \multicolumn{7}{c}{cutoff radius $r_c = 7$ \AA{}}
\end{tabular}
\end{ruledtabular}
\caption{Parameters for the model III EAM potential with $r$ in units of \AA{}
and $V(r)$ in eV.}
\end{table*}

\begin{figure*}
\includegraphics{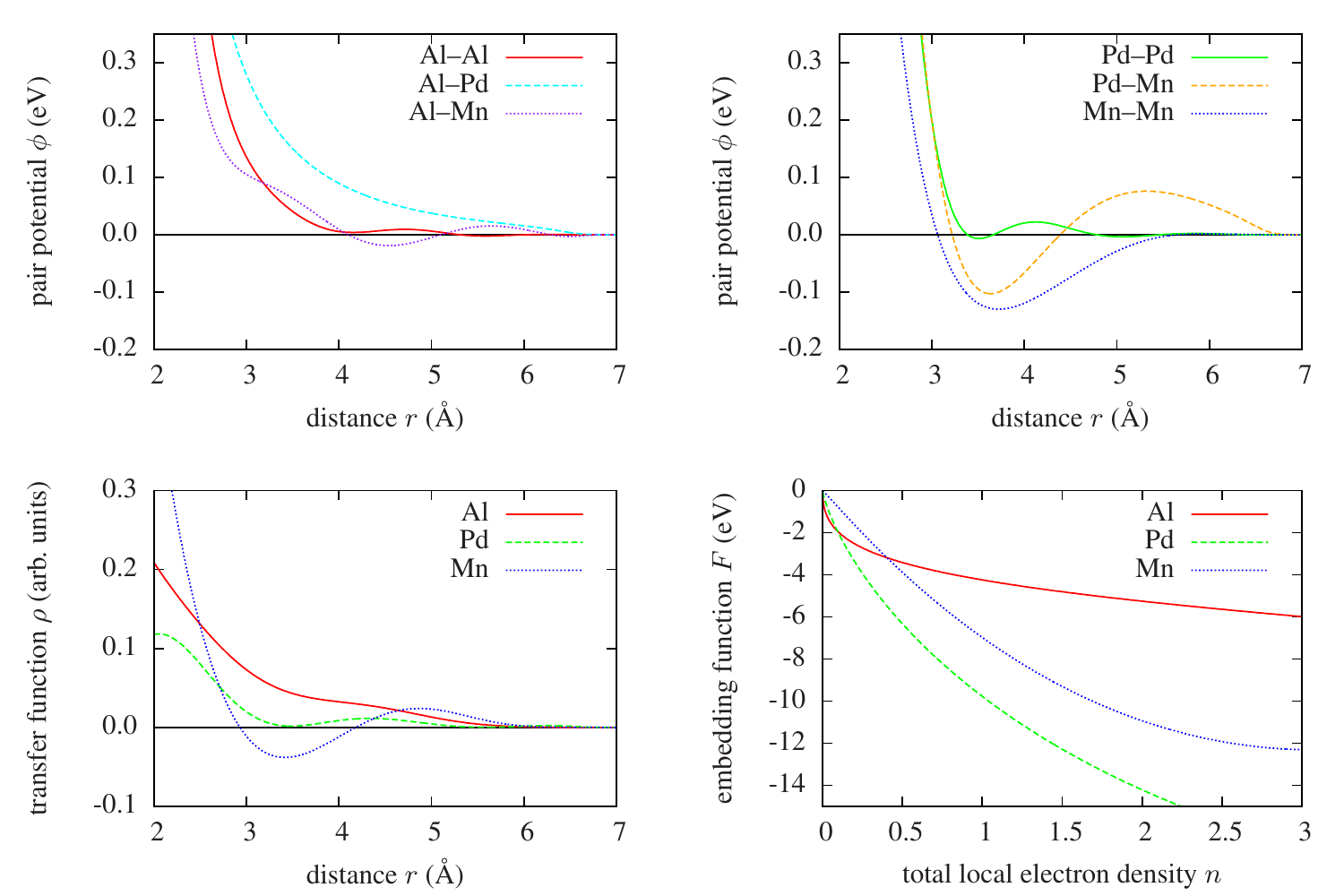}
\caption{(color online) Plots of the 12 functions of the EAM potential (model III)
for Al--Pd--Mn.}
\label{fig:potential}
\end{figure*}

The differences of the models become visible when calculating energy differences
like formation enthalpies or the convex hull. There, model III shows the smallest
deviations and can reproduce the \emph{ab initio} values with very high accuracy. The
models I and II also give very good energies differences but cannot be used to
predict the stability of a structure with the calculation of the convex hull. This
indicates that oscillations on two length scales, like in model III, are necessary.
However, the reasons for this are unclear. For further structure determination and
analysis of the metadislocations in the $\Xi$-phases, the model potential III will
be used.

\begin{acknowledgments}
We would like to thank Alejandro Santana Bonilla and Marek Mihalkovi\v{c}
for intensive discussions and providing some test and reference data. This project
has been funded by the European Network of Excellence ``Complex Metallic Alloys''
(NMP3-CT-2005-500140) and by Deutsche Forschungsgemeinschaft, Paketantrag
``Physical Properties of Complex Metallic Alloys, (PAK 36)'', TR 154/24-2.
\end{acknowledgments}

\bibliographystyle{apsrev4-1}
\bibliography{AlMnPd}

\end{document}